\documentclass{jfm}
\usepackage{graphicx}
\usepackage{epstopdf, epsfig}
\usepackage{yhmath}
\usepackage{color}
\usepackage{datetime}
\usepackage{url}

\shorttitle{Tristable flow states and reversals of the LSC}
\shortauthor{A. Xu, X. Chen and H.-D. Xi}

\title{Tristable flow states and reversal of the large-scale circulation in two-dimensional circular convection cells}

\author{
Ao Xu\aff{1},
Xin Chen\aff{1}
\and Heng-Dong Xi\aff{1}\aff{2}
\corresp{\email{hengdongxi@nwpu.edu.cn}}
}

\affiliation{
\aff{1}School of Aeronautics, Northwestern Polytechnical University, Xi'an 710072, PR China
\aff{2}Institute of Extreme Mechanics, Northwestern Polytechnical University, Xi'an 710072, PR China}

\begin{document}
\maketitle

\begin{abstract}
We present a numerical study of the flow states and reversals of the large-scale circulation (LSC) in a two-dimensional circular Rayleigh-B\'enard cell.
Long-time direct numerical simulations are carried out in the Rayleigh number ($Ra$) range $10^{7} \le Ra \le 10^{8}$ and  Prandtl number ($Pr$) range $2.0 \le Pr \le 20.0$.
We found that a new, long-lived, chaotic flow state exists, in addition to the commonly observed circulation states (the LSC in the clockwise and counterclockwise directions).
The circulation states consist of one primary roll in the middle and two secondary rolls near the top and bottom circular walls.
The primary roll becomes stronger and larger, while the two secondary rolls diminish, with increasing $Ra$.
Our results suggest that the reversal of the LSC is accompanied by the secondary rolls growing, breaking the primary roll and then connecting to form a new primary roll with reversed direction.
We mapped out the phase diagram of the existence of the LSC and the reversal in the $Ra$-$Pr$ space, which reveals that the flow is in the circulation states when $Ra$ is large and $Pr$ is small.
The reversal of the LSC can only occur in a limited $Pr$ range.
The phase diagram can be understood in terms of competition between the thermal and viscous diffusions.
We also found that the internal flow states manifested themselves into global properties such as Nusselt and Reynolds numbers.
\end{abstract}

\begin{keywords}
B{\'e}nard convection, plumes/thermals, turbulent convection
\end{keywords}

\section{Introduction}\label{sec:Introduction}

Spontaneous flow reversals occur widely in astrophysical and geophysical systems, such as reversals of the convective wind in the atmosphere \citep{van2000statistics,gallet2012reversals} and reversals of the convective flow in the Earth or stars \citep{glatzmaier1999role,miesch2009turbulence,petrelis2009simple,biggin2012possible}.
A paradigm for studying the mechanism of the flow reversals is the Rayleigh-B\'enard convection, which is a fluid layer heated from the bottom and cooled from the top \citep{ahlers2009heat,lohse2010small,chilla2012new,xia2013current}.
The control parameters of the system include the Rayleigh number $Ra = \alpha g \Delta_{T}H^{3}/(\nu \kappa)$, which describes the strength of buoyancy force relative to thermal and viscous dissipative effects; and the Prandtl number $Pr = \nu/\kappa$, which describes the thermophysical fluid properties.
Here, $\alpha$, $\kappa$ and $\nu$ are the thermal expansion coefficient, thermal diffusivity and kinematic viscosity of the fluid, respectively,
$g$ is the gravitational acceleration and $\Delta_{T}=T_{hot}-T_{cold}$ is the temperature difference across the fluid layer of height $H$. At sufficiently high $Ra$, a large-scale circulation (LSC) that spans the size of the convection cell is established \citep{krishnamurti1981large,cioni1997strongly}, which essentially results from the organization of thermal plume motion due to plume-vortex and plume-plume interactions \citep{xi2004laminar,zhou2010physical}.
An interesting feature of the LSC is the spontaneous and random reversal of its flow direction
\citep{sreenivasan2002mean,brown2005reorientation,brown2006rotations,xi2007cessations,xi2008azimuthal,Xiao2018PRE}.
To minimize or even eliminate the influence from the three-dimensional complex dynamic features of the LSC in cylindrical cells \citep{funfschilling2004plume,xi2009origin} and in cubic cells \citep{kaczorowski2013jfm}, a two-dimensional (2-D) or quasi-2-D square cell was used to study the reversal of the LSC \citep{xia2003particle,sugiyama2010flow,wagner2013aspect}, in which the LSC is confined within the 2-D or quasi-2-D plane, while the flow switches between a clockwise circulation state and a counterclockwise circulation state.
During the reversal process, the corner rolls play a crucial role  \citep{sugiyama2010flow}.
They grow in size and kinetic energy to weaken and eventually break up the primary circulating roll.
After that, two corner rolls connect to form a new primary roll again, but with a reversed circulating direction \citep{sugiyama2010flow,chandra2013flow}.
Recently, it was shown that in quasi-2-D square cells, the corner roll is not necessary for the occurrence of reversals.
The probability of the occurrence of reversals mainly depends on the stability of the LSC itself.
The role of the corner roll is to destabilize the LSC, thus increasing the reversal frequency \citep{chen2019emergence,chen2020reduced}.

To eliminate the effect of corner rolls on the reversals of the LSC, \cite{wang2018mechanism} conducted experiments in a quasi-2-D cell with its circular plane orientated parallel to gravity (referred to as a vertical circular thin disc from now on).
The top one-third and bottom one-third of the circular sidewalls are kept at low and high temperatures, respectively; the left and right sidewalls, each being one-sixth of the whole circular wall, are made of thermally insulating Plexiglas \citep{song2011coherent,song2014dynamics,wang2018mechanism,wang2018boundary}.
The shape of such a specially designed cell matches the single-roll structure of the LSC perfectly; thus, there are no corner rolls to destabilize the LSC.
Through visualization of the LSC and plume dynamics, \cite{wang2018mechanism} proposed that the reversal of the LSC is due to the rare massive eruption of thermal plumes disrupting the existing LSC and resetting its rotational direction.
The massive eruption was attributed to boundary layer instability induced by continuous heat accumulation or spontaneous fluctuations in turbulent thermal convection \citep{wang2018mechanism}. However, a general understanding of the reversal mechanism remains elusive.

In this paper, inspired by the experimental study with the vertical circular thin disc of \cite{wang2018mechanism}, we present a numerical study of the flow topology and reversal of the LSC in a 2-D circular cell with its circular plane orientated parallel to gravity.
Long-time direct numerical simulations with at least 100 000 free-fall time units were performed in the Rayleigh number range $10^{7} \le Ra \le 10^{8}$ and Prandtl number range $2.0 \le Pr \le 20.0$.
We found that in addition to the commonly observed circulation state (namely the LSC), the flow spends a considerable time in a chaotic state where the plumes go up and down randomly.
This newly observed chaotic state prevails at low $Ra$ while the circulation state prevails at high $Ra$. We also mapped out the phase diagram of the existence of the LSC and its reversal on the $Ra$-$Pr$ plane. The rest of this paper is organized as follows.
In Section \ref{sec:numericalMethod}, we present the numerical details for the simulations.
In Section \ref{sec:resultsAndDiscussion}, general flow and heat transfer features in the 2-D circular cell are presented, and the dynamics of the novel tristable flow states in the circular cell are introduced;
after that, Fourier mode decomposition analysis is adopted to understand the effects of $Ra$ and $Pr$ on the flow states.
In Section \ref{sec:conclusions}, the main findings of the present work are summarized.

\section{Numerical method}\label{sec:numericalMethod}
\subsection{Direct numerical simulation of incompressible thermal flows}
We consider incompressible thermal flows under the Boussinesq approximation: the temperature is treated as an active scalar, and its influence on the velocity field is realized through the buoyancy term; all the transport coefficients are assumed to be constants.
The governing equations can be written as
\begin{equation}
\nabla \cdot \mathbf{u} = 0 \label{Eq:continuity}
\end{equation}
\begin{equation}
\frac{\partial \mathbf{u}}{\partial t}+\mathbf{u}\cdot \nabla \mathbf{u}
=-\frac{1}{\rho_{0}}\nabla p+\nu\nabla^{2}\mathbf{u}+g\alpha(T-T_{0})\hat{\mathbf{y}} \label{Eq:momentum}
\end{equation}
\begin{equation}
\frac{\partial T}{\partial t}+\mathbf{u}\cdot\nabla T=\kappa \nabla^{2}T \label{Eq:temperatureCDE}
\end{equation}
where $\mathbf{u}=(u,v)$  is the fluid velocity,
$p$ and $T$ are pressure and temperature of the fluid, respectively.
$\rho_{0}$  and $T_{0}$  are reference density and temperature, respectively,
and $\hat{\mathbf{y}}$ is the unit vector in the vertical direction.

We adopt the lattice Boltzmann (LB) method as the numerical tool to solve the above equations.
The advantages of the LB method include easy implementation and parallelization as well as low numerical dissipation \citep{chen1998lattice,aidun2010lattice,xu2017lattice}.
In this work, the LB method is based on a double distribution function approach, which consists of a D2Q9 discrete velocity model for the Navier-Stokes equations to simulate fluid flows (i.e. solving Eqs. \ref{Eq:continuity} and \ref{Eq:momentum}) and a D2Q5 discrete velocity model for the convection-diffusion equation to simulate heat transfer (i.e. solving Eq. \ref{Eq:temperatureCDE}).
To simulate fluid flows and heat transfer in the circular domain, we apply the interpolated bounce-back scheme at the curved wall.
Specifically, to guarantee no-slip velocity boundary conditions at the curved wall, we apply the interpolated bounce-back scheme of  \citet{bouzidi2001momentum} for the density distribution function;
to achieve the Dirichlet or Neumann temperature boundary conditions at the curved wall, we apply the interpolated bounce-back scheme of  \citet{li2013boundary} for the temperature distribution function.
More numerical details of the LB method and validation of the in-house code can be found in our previous work \citep{xu2017accelerated,xu2019lattice,xu2019statistics}.

\subsection{Simulation settings}

\begin{figure}
  \centerline{\includegraphics[width=5cm]{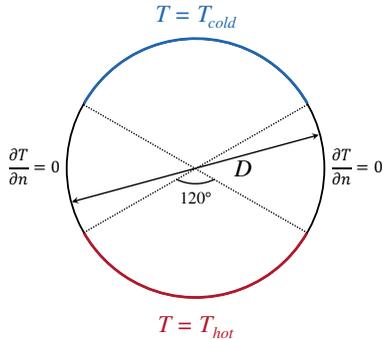}}
  \caption{Schematic illustration of the 2-D circular convection cell.}
\label{fig:cellDemo}
\end{figure}

As illustrated in figure \ref{fig:cellDemo}, the top (blue curve) and the bottom (red curve) one-third of the circular walls are kept at constant low and high temperature of $T_{cold}$ and $T_{hot}$, respectively, while the left and right one-sixth of the circular sidewalls (black curves) are adiabatic.
All circular walls impose no-slip velocity boundary conditions.
The diameter of the circular cell is $D$ and is chosen as the characteristic length of the system.
Simulation results are provided in the Rayleigh number range $10^{7} \le Ra \le 10^{8}$ and Prandtl number range $2.0 \le Pr \le 20.0$.
To have better statistics, we run each simulation for at least 100 000 $t_{f}$,  and in some cases the simulation times are as long as 200 000 $t_{f}$.
Here, $t_{f}$ denotes free-fall time units: $t_{f}=\sqrt{D/(g\alpha\Delta_{T})}$.
We also check whether the grid spacing $\Delta_{g}$ and time interval $\Delta_{t}$ are adequately resolved by comparing with the Kolmogorov and Batchelor scales.
Here, we roughly estimate the global kinetic energy dissipation rate as $\langle \varepsilon_{u} \rangle=RaPr^{-2}(Nu-1)\nu^{3}/D^{4}$  \citep{shraiman1990heat}.
Thus, the Kolmogorov length scale can be estimated by the global relation as $\eta_{K}=(\nu^{3}/\langle \varepsilon_{u} \rangle)^{1/4}=DPr^{1/2}/[Ra(Nu-1)]^{1/4}$, the Batchelor length scale can be estimated as $\eta_{B}=\eta_{K}Pr^{-1/2}$  \citep{silano2010numerical} and the Kolmogorov time scale can be estimated as $\tau_{\eta}=\sqrt{\nu/\langle\varepsilon_{u}  \rangle }=t_{f}\sqrt{Pr/(Nu-1)}$.
The global heat transport is measured by the wall-averaged Nusselt number as $Nu=-(\langle \partial_{\mathbf{n}}T \rangle_{top,t}+\langle \partial_{\mathbf{n}}T \rangle_{bottom,t})/2$, where $\langle \cdots \rangle_{top/bottom,t}$  denotes the ensemble average over the top (or bottom) one-third of the circular walls and over the time \citep{kerr1996rayleigh,verzicco2003numerical} and $\mathbf{n}$ denotes the unit vector in the wall norm direction.
From table \ref{tab:resolution}, we can see that grid spacing satisfies the criterion of $\max(\Delta_{g}/\eta_{K},\Delta_{g}/\eta_{B})\le0.47$, which ensures spatial resolution.
Besides, the time intervals are $\Delta_{t}\le 0.0095\tau_{\eta}$; thus, adequate temporal resolution is guaranteed.

\begin{table}
  \begin{center}
\def~{\hphantom{0}}
  \begin{tabular}{ccccccccccccccccccccccccccccccccccccccc}
  $Ra$ & $Pr$ & $D$ & $\Delta_{g}/\eta_{K}$ & $\Delta_{g}/\eta_{B}$ & $\Delta_{t}/\tau_{\eta}\times 10^{3}$ \\ [3pt]
  $1.0\times 10^{7}$   & [2.0, 20.0] & 257 & [0.10, 0.29] & [0.41, 0.43] & [3.4, 9.5] \\
  $2.0\times 10^{7}$   & [2.0, 20.0] & 385 & [0.08, 0.24] & [0.34, 0.36] & [2.5, 7.0] \\
  $3.0\times 10^{7}$   & [2.0, 20.0] & 385 & [0.09, 0.27] & [0.38, 0.41] & [2.7, 7.3] \\
  $4.5\times 10^{7}$   & [2.0, 20.0] & 385 & [0.10, 0.31] & [0.44, 0.47] & [2.8, 7.9] \\
  $6.0\times 10^{7}$   & [2.0, 20.0] & 513 & [0.09, 0.26] & [0.36, 0.38] & [2.2, 6.2] \\
  $8.0\times 10^{7}$   & [2.0, 20.0] & 513 & [0.09, 0.28] & [0.40, 0.42] & [2.3, 6.6] \\
  $1.0\times 10^{8}$   & [2.0, 20.0] & 513 & [0.10, 0.31] & [0.43, 0.45] & [2.4, 6.9] \\
  \end{tabular}
  \caption{Spatial and temporal resolutions of the simulations.}
  \label{tab:resolution}
  \end{center}
\end{table}

\section{Results and discussion}\label{sec:resultsAndDiscussion}

\subsection{Global flow and heat transfer features}

\begin{figure}
  \centerline{\includegraphics[width=12cm]{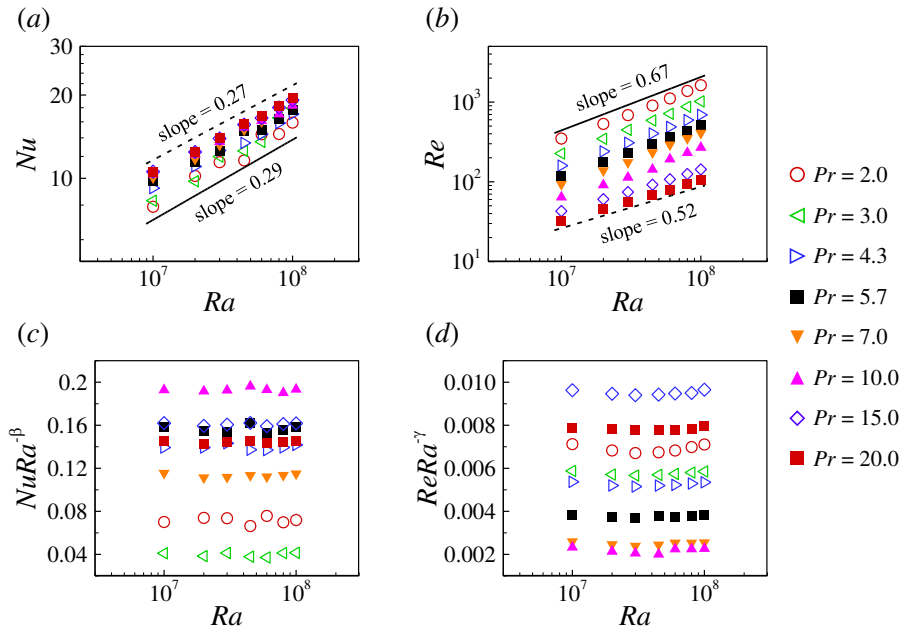}}
  \caption{(\textit{a}) Nusselt number, (\textit{b}) Reynolds number, (\textit{c}) compensated Nusselt number and (\textit{d}) compensated Reynolds number as a function of Rayleigh number for various Prandtl numbers.
  The power-law exponents $\beta$ and $\gamma$ used in the compensated plots are, respectively:
  0.29 and 0.67 for $Pr=2.0$; 0.33 and 0.66 for $Pr=3.0$; 0.26 and 0.64 for $Pr=4.3$;
  0.26 and 0.64 for $Pr=5.7$; 0.28 and 0.65 for $Pr=7.0$; 0.25 and 0.63 for $Pr=10.0$;
  0.26 and 0.52 for $Pr=15.0$; 0.27 and 0.52 for $Pr=20.0$.}
\label{fig:Nu_Re}
\end{figure}

We first examine the scaling of the global quantities, such as Nusselt number ($Nu$) and Reynolds number ($Re$), on the control parameter $Ra$.
The measured $Nu$ and $Re$ as a function of $Ra$ for various $Pr$ are shown in figures \ref{fig:Nu_Re}(a) and \ref{fig:Nu_Re}(b), respectively.
Due to the zero value of the averaged velocity over the whole cell, we choose the root-mean-square velocity $u_{rms}=\sqrt{\langle  (u^{2}+v^{2}) \rangle_{V,t}}$  as the characteristic velocity to calculate the Reynolds number $Re=u_{rms}D/\nu$ \citep{sugiyama2009flow}.
Here, $\langle \cdots \rangle_{V,t}$ denotes the volume and time average.
The data shown in figure \ref{fig:Nu_Re} can be described by power-law relations for each $Pr$, namely $Nu \propto Ra^{\beta}$ and $Re \propto Ra^{\gamma}$.
To further confirm that the data faithfully follow the power law, we show the compensated plots of $Nu$ and $Re$ as a function of $Ra$ in figures \ref{fig:Nu_Re}(c) and \ref{fig:Nu_Re}(d), respectively.
The heat transfer scaling exponent $\beta$ ranges between 0.25 and 0.33 for various $Pr$, while the momentum scaling exponent $\gamma$ ranges between 0.52 and 0.67.
The $Nu(Ra)$ and $Re(Ra)$ scaling exponents in the 2-D circular cell are close to those in a 2-D (or quasi-2-D) square cell, where $\beta = 0.285 \sim 0.30$ \citep{ciliberto1996large,van2012flow,huang2016effects,zhang2017statistics}  and $\gamma = 0.59 \sim 0.62$ \citep{xia2003particle,sun2005scaling,zhou2007measured,van2012flow,zhang2017statistics}  for moderate $Pr$.
In addition, the heat transfer scaling exponent $\beta$  from our simulations ($\beta = 0.26$ for $Pr = 4.3$ and $\beta = 0.25$ for $Pr = 5.7$) is very close to experimental measurements \citep{song2011coherent,song2014dynamics,wang2018mechanism,wang2018boundary} in the quasi-2-D circular cell ($\beta=0.275$ for $Pr = 4.4$ and $\beta=0.28$ for $Pr = 5.4$).
The differences in the momentum scaling exponent $\gamma$, namely $\gamma = 0.64$ for $Pr = 5.7$ from our simulations compared with $\gamma=0.49$ for $Pr = 5.4$ from experimental measurements \citep{song2011coherent,song2014dynamics,wang2018mechanism,wang2018boundary}, are due to stronger convection in two-dimensional cells than that in three-dimensional cells \citep{van2013comparison}.
Overall, the global heat transfer and momentum quantities reveal that the simulated system possesses the key features of turbulent convection.

\subsection{The tristable flow states and reversals of the LSC \label{sec:tristable}}

\begin{figure}
  \centerline{\includegraphics[width=10cm]{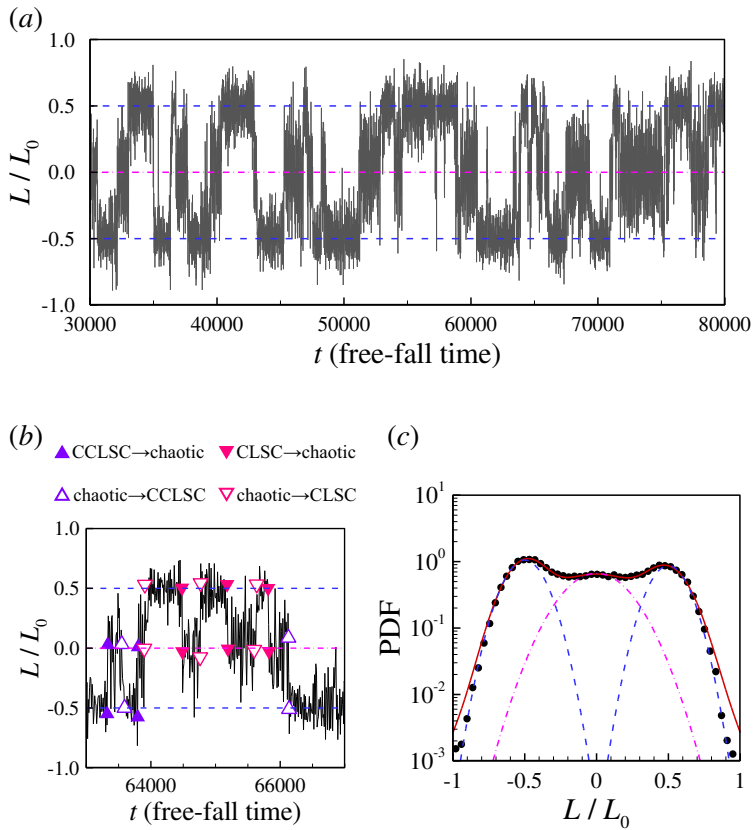}}
  \caption{
  (\textit{a}) Time trace of the global angular momentum $L$ of the flow for $Ra = 10^{7}$ and $Pr = 4.3$ (non-dimensionalized by its maximum value $L_{0}$).
  (\textit{b}) Enlarged portion of the data shown in (\textit{a}) to illustrate the tristable states.
  (\textit{c}) The PDF of the dimensionless global angular momentum. The three dashed lines are the fitting of the data to three independent Gaussian functions, while the solid line is the fitting of the data to a triple-Gaussian function.}
\label{fig:L_trace}
\end{figure}

To study the time evolution of the flow topology, we calculate the global angular momentum $L(t)=\langle -(y-D/2)u(\mathbf{x},t)+(x-D/2)v(\mathbf{x},t)\rangle_{V}$  of the flow field, where $\langle \cdots \rangle_{V}$ denotes the volume average.
Previously, it was shown that the global angular momentum could nicely indicate different flow states in a 2-D circular cell \citep{sugiyama2010flow}.
Figure \ref{fig:L_trace}(a) shows the time series of the dimensionless global angular momentum $L(t)/L_{0}$ at $Ra = 10^{7}$ and $Pr = 4.3$ for the period of 50 000 free-fall time units.
Here, $L_{0}$ is the maximum of the absolute value of $L(t)$.
We can see from figure \ref{fig:L_trace}(a) that $L$ switches between three meta-stable states: one state with positive fluctuating $L$ well above zero, which corresponds to the LSC in the clockwise direction (referred to as CLSC); one state with negative fluctuating $L$ well below zero, which corresponds to the LSC in the counterclockwise direction (referred to as CCLSC); and one state with $L$ fluctuating around zero, which is referred to as chaotic state with random motion of plumes.
The tristable flow states are much clearer in figure \ref{fig:L_trace}(b), where an enlarged view of the time segments of $L(t)/L_{0}$ is shown.
We use the following four pairs of symbols to denote the flow state transitions: (i) up filled triangles mark the transitions from the CCLSC to chaotic states, (ii) up open triangles mark the transitions from the chaotic to CCLSC states, (iii) down filled triangles mark the transitions from the CLSC to chaotic states and (iv) down open triangles mark the transitions from the chaotic to CLSC states.
The tristable behaviour of the flow state can also be seen from the probability density functions (PDFs) of $L(t)/L_{0}$ shown in figure \ref{fig:L_trace}(c). The three dashed lines are the fitting results using three independent Gaussian functions, while the solid line is the fitting result using a triple-Gaussian function.
The most probable values of $L(t)/L_{0}$ for these three states are around $\pm 0.5$ and 0, respectively.
This tristable flow state feature is distinguishable from the bistable behaviour in 2-D or quasi-2-D square cells, where the LSC switches only between CLSC and CCLSC, and a double-Gaussian function best fits the PDFs of the flow strength in terms of either global angular momentum or the temperature contrast \citep{sugiyama2010flow, chandra2011dynamics,ni2015reversals,huang2016effects,castillo2016reversal,castillo2019cessation}.

\begin{figure}
  \centerline{\includegraphics[width=9cm]{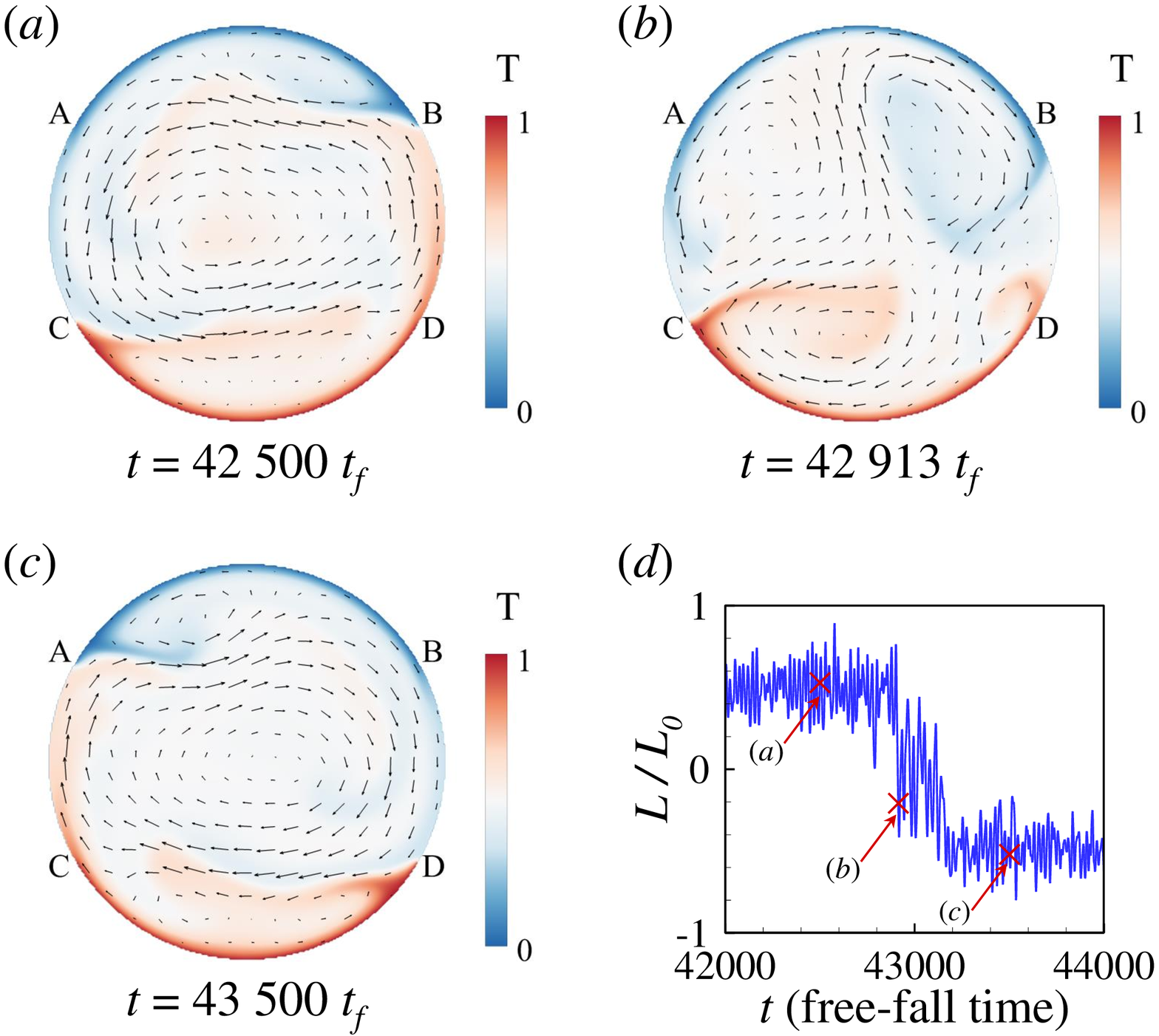}}
  \caption{Typical snapshots of temperature field (contours) and velocity field (arrows) of the flow field at (\textit{a}) $t = 42 \ 500 \ t_{f}$, (\textit{b}) $t = 42 \ 913 \ t_{f}$ and (\textit{c}) $t = 43 \ 500 \ t_{f}$. (\textit{d}) Time segment of the dimensionless global angular momentum $L(t)/L_{0}$.  The flow fields and $L(t)/L_{0}$ are obtained at $Ra = 10^{7}$ and $Pr = 4.3$. Arcs $\wideparen{AB}$ and $\wideparen{CD}$ denote the top and bottom one-third of the circular walls, respectively.}
\label{fig:three_states}
\end{figure}

Typical snapshots of temperature and velocity fields for the three meta-stable flow states at $Ra = 10^{7}$ and $Pr = 4.3$ are shown in figures \ref{fig:three_states}(a-c), and the corresponding movie can be viewed as supplementary movie 1 available at \url{https://doi.org/10.1017/jfm.2020.964}. From the movie and snapshots, we can identify the two circulation states, namely both the CCLSC and CLSC (see figures \ref{fig:three_states}a and \ref{fig:three_states}c, respectively);
between the two circulation states is the chaotic state, where thermal plumes erupt randomly and fail to be organized (see figure \ref{fig:three_states}b).
Figure \ref{fig:three_states}(d) further shows the time segments of the dimensionless angular momentum $L(t)/L_{0}$ and the value of $L(t)/L_{0}$ of the three flow states shown in figure \ref{fig:three_states}(a-c).
It should be noted that the chaotic state in our study is different from the short transient state (i.e. the flow switching between the CLSC and CCLSC) in 2-D or quasi-2-D square cells, as the duration of the former is two or three orders of magnitude longer than that of the latter.

\begin{figure}
  \centerline{\includegraphics[width=11cm]{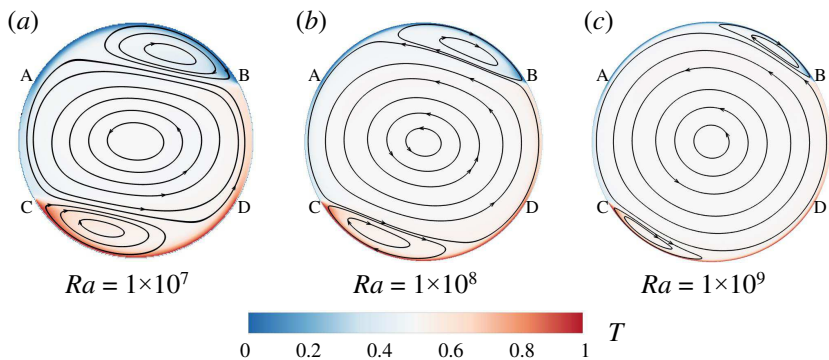}}
  \caption{Time-averaged temperature field (contours) and velocity field (arrows) of the flow for (\textit{a}) $Ra =1\times 10^{7}$ ,  (\textit{b}) $Ra = 1\times 10^{8}$ and  (\textit{c}) $Ra = 1\times10^9$ and $Pr = 4.3$. Arcs $\wideparen{AB}$ and $\wideparen{CD}$ denote the top and bottom one-third of the circular walls, respectively.}
\label{fig:meanfield_Ra}
\end{figure}

When the flow is in the circulation state, secondary rolls develop near the top and bottom circular walls.
The formation of the secondary rolls is of the same origin as the corner rolls in (quasi-)2-D square cells:
the interaction between the descending (ascending) side of the primary roll and the ascending (descending) plumes generated from the edge of the bottom (top) wall.
Taking CCLSC as an example (see figure \ref{fig:three_states}a), the descending side of the primary roll shears the ascending plumes from the left edge of the bottom wall, such that these plumes are forced to move horizontally to the right.
Some of these plumes are deflected down when they reach the right-hand side of the wall and then follow the curved wall to the left, thus forming a small secondary roll at the bottom of the cell.
Similarly, a secondary roll is formed at the top of the cell.
These two secondary rolls are competing with the primary roll.
When $Ra$ is large, the primary roll becomes stronger, and thus it can penetrate to very close to the top and bottom circular walls, and the secondary rolls would become smaller or even disappear.
When $Ra$ is small, the primary roll becomes weaker and it cannot penetrate too much to very close to the top and bottom circular walls, and thus the secondary roll would become larger.
This is indeed the case, as shown in figure \ref{fig:meanfield_Ra}(a-c), where the long time-averaged velocity fields for $Ra = 1\times10^7$,  $ 1\times10^8$ and $1\times10^9$  and $Pr = 4.3$ are given.
We can see that with increasing $Ra$, the primary roll becomes larger, while the top and bottom secondary rolls become smaller.
We would expect that at very high $Ra$, the secondary rolls will disappear, and a single roll structure occupies the whole cell.
The above conjecture is supported by the experimentally observed single roll structure in the vertical circular thin disc at high $Ra$  ($3\times 10^{9} \le Ra \le 2\times 10^{10}$) where the secondary rolls are absent \citep{song2011coherent,song2014dynamics,wang2018mechanism,wang2018boundary}.
In addition, previous studies \citep{xia2003particle,zhou2018similarity} suggest that in (quasi-)2-D square cells, the corner rolls will also become smaller with increasing $Ra$.
In our 2-D simulations, we choose a smaller parameter space of $Ra$ compared with those in experiments \citep{song2011coherent,song2014dynamics,wang2018mechanism,wang2018boundary}.
The reason is that flow reversal starts to occur at smaller $Ra$ in the 2-D cell compared with the quasi-2-D cell \citep{sugiyama2010flow}.

\begin{figure}
  \centerline{\includegraphics[width=12cm]{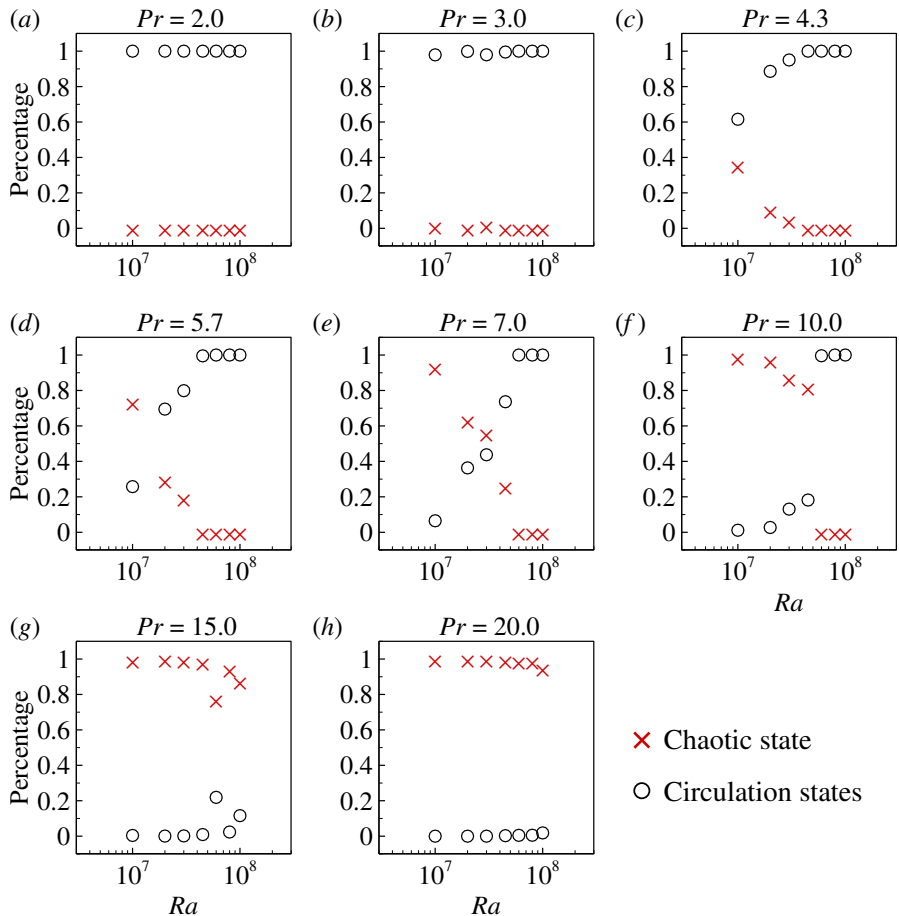}}
  \caption{The residence time, i.e. percentage of time the flow spends in the two circulation states and the chaotic state, for (\textit{a}) $Pr = 2.0$, (\textit{b}) $Pr = 3.0$, (\textit{c}) $Pr = 4.3$, (\textit{d}) $Pr = 5.7$, (\textit{e}) $Pr = 7.0$, (\textit{f}) $Pr = 10.0$, (\textit{g}) $Pr = 15.0$ and (\textit{h}) $Pr = 20.0$.}
\label{fig:residence_time}
\end{figure}

We then measure the residence time, i.e. the percentage of time the flow spends in the two circulation states and the chaotic state.
The criteria to quantitatively identify these flow state transitions are similar to those used in \cite{huang2016effects}.
First, the flow should stay at the current state for more than one large-eddy turnover time $t_{E}=4\pi/\langle |\omega_{c}(t)| \rangle$, where $\omega_{c}$ denotes the vorticity at the cell centre.
Second, the flow states should be clearly distinguished from each other, which is quantified by the three distinguishable peaks of the PDFs of $L(t)/L_{0}$.
Third, when the flow changes from one state to another, the value of $L(t)/L_{0}$ should be larger (or smaller) than the peak value in the PDF corresponding to the new flow state.
We plot the residence time as a function of $Ra$ for different $Pr$ in figure \ref{fig:residence_time}.
It is seen from the figure that for low $Pr$ (e.g.  $Pr=2.0$ and $Pr=3.0$), most of the time the flow is in the circulation states for the whole $Ra$ range investigated ($10^{7} \le Ra \le 10^{8}$).
For intermediate $Pr$ (e.g.  $4.3 \le Pr \le 10.0$), the time the flow spends in the circulation state increases with $Ra$, while the time the flow spends in the chaotic state decreases, which implies that the flow is more organized at higher $Ra$.
The fact that the circulation states are prevailing for higher $Ra$ (at intermediate $Pr$) is consistent with the experimental results for a vertical circular thin disc \citep{wang2018mechanism} where the chaotic state seems absent.
For high $Pr$ (e.g.  $Pr=15.0$ and $20.0$), most of the time, the flow is in the chaotic state for the whole $Ra$ range investigated. The $Pr$ dependency of the flow state can be understood as follows: at low $Pr$, due to the large thermal diffusivity, very energetic vertical plumes which could destroy the LSC are not likely to be formed, and thus the LSC remains stable; at higher $Pr$, due to the lower thermal diffusivity, the plumes are very likely to be energetic, they tend to ascend and descend vertically and thus fail to organize into an LSC.

\begin{figure}
  \centerline{\includegraphics[width=9cm]{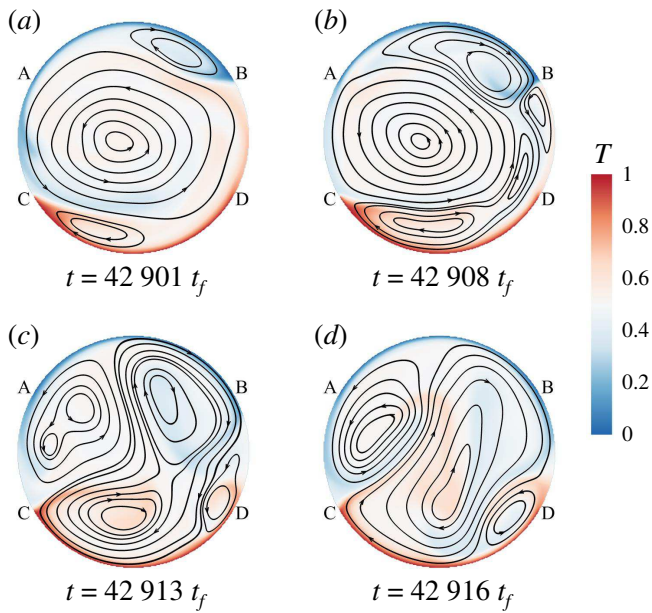}}
  \caption{Typical snapshots of the flow field (temperature field and streamlines) during the breakup of the primary roll at (\textit{a}) $t = 42 \ 901 \ t_{f}$, (\textit{b}) $t = 42 \ 908 \ t_{f}$, (\textit{c}) $t = 42 \ 913 \ t_{f}$ and (\textit{d}) $t = 42 \ 916 \ t_{f}$. Arcs $\wideparen{AB}$ and $\wideparen{CD}$ denote the top and bottom one-third of the circular walls, respectively.}
\label{fig:breaking}
\end{figure}

To understand how the flow switches from the circulation state to the chaotic state, we check the flow and temperature fields during the breakup of the primary roll, namely the transition from one of the circulation states to the chaotic state.
The breakup process of the primary roll in the circular cell is very similar to that in a 2-D square cell, as shown in figure \ref{fig:breaking}.
Detached hot (or cold) plumes from the bottom (or top) circular walls are trapped in the secondary rolls, further feeding energy to the growth of the secondary flows (see figure \ref{fig:breaking}b).
With the growth in size of the secondary rolls, they gradually break the primary roll and connect to form a new primary roll (see figure \ref{fig:breaking}c,d).
However, the newly formed roll is unstable, and it will be squeezed and further broken down by the growth of secondary rolls, and thus the flow enters a chaotic state.
This stage is different from that in a 2-D square cell, where a new LSC in the opposite direction will be organized shortly after the destruction of the old LSC, as discussed in detail in Section \ref{sec:onset}.

\begin{figure}
  \centerline{\includegraphics[width=12cm]{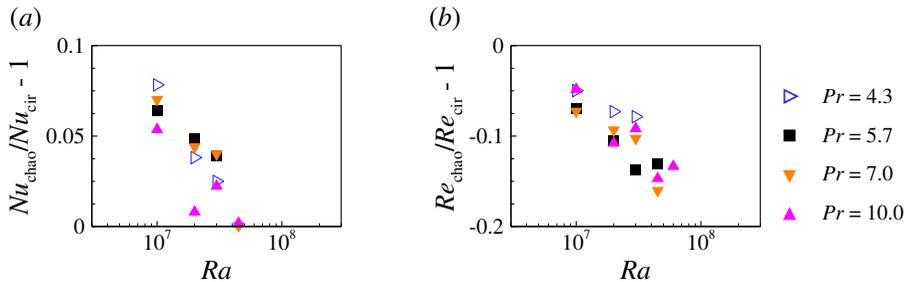}}
  \caption{Values of (\textit{a}) $Nu_{\text{chao}}/Nu_{\text{cir}}-1$  and  (\textit{b}) $Re_{\text{chao}}/Re_{\text{cir}}-1$  as functions of $Ra$ for various $Pr$. Here, $Nu_{\text{chao}}$ and $Nu_{\text{cir}}$ are the conditional averaged $Nu$ when the flow is in the chaotic state and the circulation states, respectively; $Re_{\text{chao}}$ and $Re_{\text{cir}}$ are analogously defined.}
\label{fig:Nu_diff_state}
\end{figure}

It was conjectured \citep{roche2002prandtl,chilla2004long} and shown by experiments \citep{sun2005azimuthal,xi2008flow,weiss2011large,weiss2011turbulent,xi2016higher} as well as simulations \citep{van2011connecting,van2012flow} that the internal flow states are manifested by global quantities, such as the global heat transfer.
To check whether this is the case in 2-D circular cells, we now examine the conditional averaged Nusselt and Reynolds numbers in the chaotic state and the circulation states.
In figure \ref{fig:Nu_diff_state}, we plot their differences $Nu_{\text{chao}}/Nu_{\text{cir}}-1$  and  $Re_{\text{chao}}/Re_{\text{cir}}-1$ as functions of $Ra$ for various $Pr$.
Here, the subscripts 'chao' and 'cir' denote the chaotic state and the circulation states, respectively.
Because the chaotic state is absent (or almost absent) at $Pr = 2.0$ and $3.0$, and the circulation states are absent (or almost absent) at $Pr = 15.0$ and $20.0 $, we only present results for $4.3 \le Pr \le 10.0$.
From figure \ref{fig:Nu_diff_state}, we can see that for the chaotic state, the conditional averaged $Nu$ is larger while the conditional averaged $Re$ is smaller than when the flow is in the circulation states.
The lower heat transfer efficiency in the circulation states is presumably due to the primary roll separating the top and the bottom secondary rolls, and the lack of direct communication between the top and the bottom walls results in the reduction of the heat transfer.
While in the chaotic state, the plumes erupted from the top and bottom circular walls can travel directly to the opposite walls, thus building up short channels for heat transfer, resulting in a higher $Nu$.
The behaviour of $Re$ follows an opposite trend: overall $Re$ is larger when the flow is in the circulation states, implying a stronger flow in this flow state as it is well organized.

\subsection{The onset of the LSC in circular and square cells \label{sec:onset}}

\begin{figure}
  \centerline{\includegraphics[width=12cm]{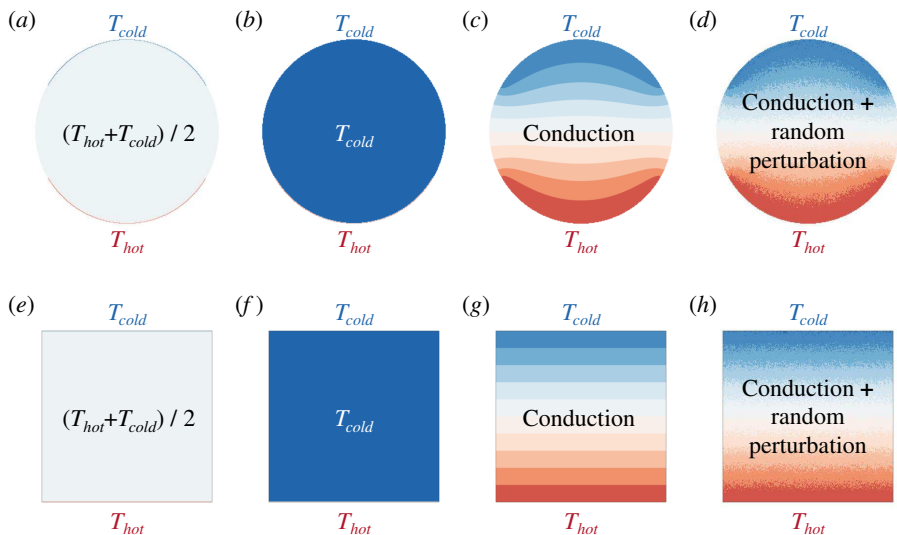}}
  \caption{Different initial settings of the temperature profiles in (\textit{a}-\textit{d}) the circular cell and (\textit{e}-\textit{h}) the square cell. (\textit{a},\textit{e}) Uniform distribution of temperature $(T_{hot}+T_{cold})/2$ in the bulk; (\textit{b},\textit{f}) uniform distribution of temperature $T_{cold}$ in the bulk; (\textit{c},\textit{g}) conduction temperature profile; (\textit{d},\textit{h}) conduction temperature profile with random perturbation.}
\label{fig:initialCondition}
\end{figure}

To understand the origin of the chaotic flow state in the 2-D circular cell, which was not observed in the 2-D square cell \citep{sugiyama2010flow}, we examine the onset process of the LSC in the circular and square cells.
Because the behaviour of the LSC onset is sensitive to initial conditions, in figure \ref{fig:initialCondition} we show four groups of different initial settings for the temperature profiles in the circular and the square cells:
(i) a uniform distribution of temperature $(T_{hot}+T_{cold})/2$ in the bulk;
(ii) a uniform distribution of temperature $T_{cold}$ in the bulk;
(iii) conduction temperature profile;
and (iv) conduction temperature profile with random perturbation.
Here, the conduction temperature profile was obtained via the steady temperature field at $Ra = 10^{3}$ and $Pr = 4.3$ (i.e. $Ra$ is below the threshold value for the onset of thermal convection).
The random perturbation was achieved via superposing a uniformly distributed random number with a maximum/minimum value of $\pm 0.05(T_{hot}-T_{cold})$ to the local temperature field.

\begin{figure}
  \centerline{\includegraphics[width=10cm]{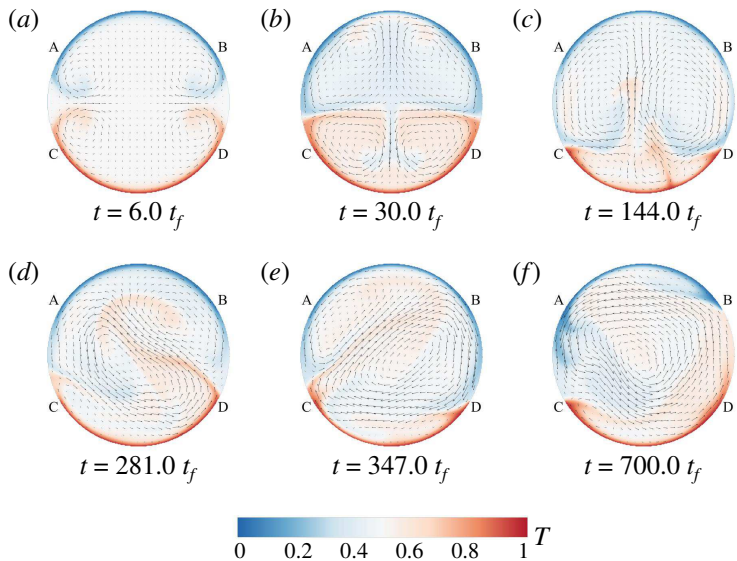}}
  \caption{Snapshots of the flow fields (both temperature contours and velocity vectors) showing the onset of LSC in the 2-D circular cell at (\textit{a}) $t = 6.0 \ t_{f}$, (\textit{b}) $t = 30.0 \ t_{f}$, (\textit{c}) $t = 144.0 \ t_{f}$, (\textit{d}) $t = 281.0 \ t_{f}$, (\textit{e}) $t = 347.0 \ t_{f}$ and (\textit{f}) $t = 700.0 \ t_{f}$ for $Ra = 10^{7}$ and $Pr = 4.3$.
  Initially, a uniform distribution of temperature $(T_{hot}+T_{cold})/2$ in the bulk.
  It takes roughly $450 \ t_{f}$ for the LSC to build up. Arcs $\wideparen{AB}$ and $\wideparen{CD}$ denote the top and bottom one-third of the  circular walls, respectively.}
\label{fig:circular_onset}
\end{figure}

\begin{figure}
  \centerline{\includegraphics[width=10cm]{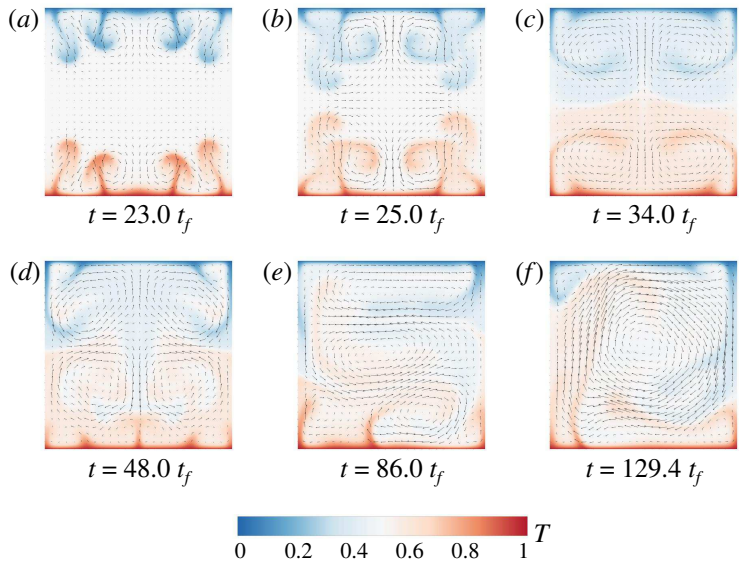}}
  \caption{Snapshots of the flow fields (both temperature contours and velocity vectors) showing the onset of LSC in the 2-D square cell at (\textit{a}) $t = 23.0 \ t_{f}$, (\textit{b}) $t = 25.0 \ t_{f}$, (\textit{c}) $t = 34.0 \ t_{f}$, (\textit{d}) $t = 48.0 \ t_{f}$, (\textit{e}) $t = 86.0 \ t_{f}$ and (\textit{f}) $t = 129.4 \ t_{f}$ for $Ra = 10^{7}$ and $Pr = 4.3$.
  Initially, a uniform distribution of temperature $(T_{hot}+T_{cold})/2$ in the bulk.
  It takes roughly $130 \ t_{f}$ for the LSC to build up.}
\label{fig:square_onset}
\end{figure}

In the following, we discuss in detail the onset of the LSC with the first group of initial temperature settings.
For the other three groups of initial temperature settings, the onset processes exhibit similar even though not identical patterns.
Figures \ref{fig:circular_onset} and \ref{fig:square_onset} show the typical snapshots of temperature and velocity fields during the onset of the LSC in circular and square cells, respectively.
The corresponding movies can be viewed from the supplementary materials (movies 2 and 3).
In the circular cell, starting from the time when the temperature difference is just applied to the top and bottom circular walls (referred to as quiescent state; see figure \ref{fig:initialCondition}a), the thermal boundary layers develop near the top and bottom circular walls, the thermal plumes grow and detach from the edge of the boundary layers (marked as points $A$, $B$, $C$ and $D$ in figure \ref{fig:circular_onset}), where the temperature boundary conditions of the wall change abruptly from constant temperature to adiabatic; the temperature field preserves both the top-bottom symmetry and left-right symmetry (see figure \ref{fig:circular_onset}a).
With the growth of plumes, the interactions between hot and cold plumes lead to the breakdown of the top-down symmetry of the temperature field, while the left-right symmetry is still kept (see figure \ref{fig:circular_onset}b).
After that, the plumes start to have wavy-like motions, and the left-right symmetry of the temperature field is also broken.
The plumes may also erupt from the centre of the boundary layers (see figure \ref{fig:circular_onset}c).
Then, the flow enters the chaotic state, where massive eruptions of the thermal plumes from either the left or the right edge of the boundary layer prevent the establishment of the LSC, and we can observe unstable clockwise or counterclockwise rotation roll in the cell centre (see figure \ref{fig:circular_onset}d,e).
The competition between the clockwise and counterclockwise rotation rolls continues until one of them prevails, and a stable LSC is built up (see figure \ref{fig:circular_onset}f).

For comparison, we also study the onset of the LSC in the 2-D square cell.
In the square cell, starting from the quiescent state (see figure \ref{fig:initialCondition}e), the thermal boundary layers develop near the top and bottom flat walls, and thermal plumes detach from random locations on the boundary layers (see figures \ref{fig:square_onset}a-d).
In the beginning, the flow is both up-down and left-right symmetric.
When $t = 48.0 \ t_f$, the up-down symmetry is broken. After the symmetry of the temperature field is broken, the plumes start to have wavy-like motions (see figure \ref{fig:square_onset}e).
With more interactions among the plumes, the plumes start to organize spatially, where hot rising plumes move along the left sidewall and cold falling plumes move along the right sidewall (see figure \ref{fig:square_onset}f).
The flow gradually evolves into a clockwise circulatory motion, and the LSC is formed.
In the square cell, the symmetry of the temperature field is more easily broken compared with that in the circular cell. Once the symmetry of the temperature field is broken, the LSC will be quickly organized.
It should be noted that the initial transient stage described here is different from that of experimental investigation \citep{xi2004laminar},
where the heating and cooling capacities for the bottom and top walls are not identical, and thus the top-down boundary conditions are slightly asymmetric.

\begin{figure}
  \centerline{\includegraphics[width=10cm]{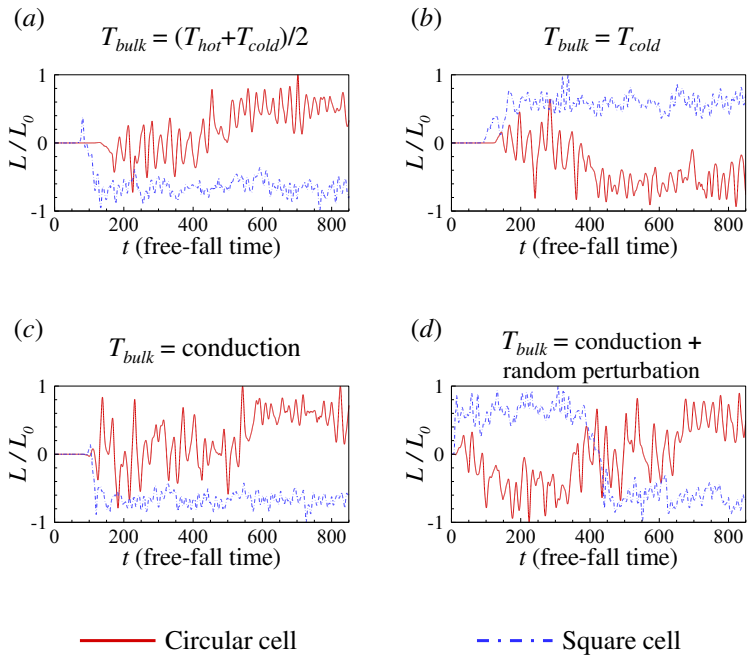}}
  \caption{The time trace of the dimensionless global angular momentum $L(t)/L_{0}$ in the 2-D circular and square cells with $Ra = 10^{7}$ and $Pr = 4.3$, showing the onset of LSC, for different initial settings: (\textit{a}) uniform distribution of temperature $(T_{hot}+T_{cold})/2$ in the bulk; (\textit{b}) uniform distribution of temperature $T_{cold}$ in the bulk; (\textit{c}) conduction temperature profile; (\textit{d}) conduction temperature profile with random perturbation.}
\label{fig:onset_L}
\end{figure}

To further compare the organization of thermal plumes into the LSC in the 2-D circular and square cells, we plot the dimensionless global angular momentum $L(t)/L_{0}$ during the onset process of the LSC, as shown in figure \ref{fig:onset_L}.
For all the four cases, in the circular cell, $L/L_{0}$ fluctuates around zero, and it takes a while (around $t \gtrsim 500 \ t_{f}$ in the first three cases and $t \gtrsim 100 \ t_{f}$ in the last case) to reach a plateau well above zero, and the LSC is organized.
In contrast, in the square cell, $L/L_{0}$ reaches a plateau well below (or above) zero with negative (or positive) fluctuations much earlier than that in the square cell (around $t \gtrsim 200 \ t_{f}$  in the first three cases and $t \gtrsim 20 \ t_{f}$  in the last case), indicating that the LSC in the clockwise (or counterclockwise) direction is organized.
Specifically, in the circular cell, the plumes detach mainly from the edge of the top and bottom circular walls (marked as points $A$, $B$, $C$ and $D$ in figure \ref{fig:circular_onset}).
In contrast, in the square cell, the plumes detach from random horizontal positions on the top and bottom circular walls (see figure \ref{fig:square_onset}), and there is no competition between rolls rotated in opposite directions.
Thus, in the 2-D circular cell, it takes a longer time to establish a stable LSC, and the obstruction to building the stable LSC may further result in the chaotic flow state.

\subsection{Flow mode decomposition and mode evolution during a reversal}

\begin{figure}
  \centerline{\includegraphics[width=11cm]{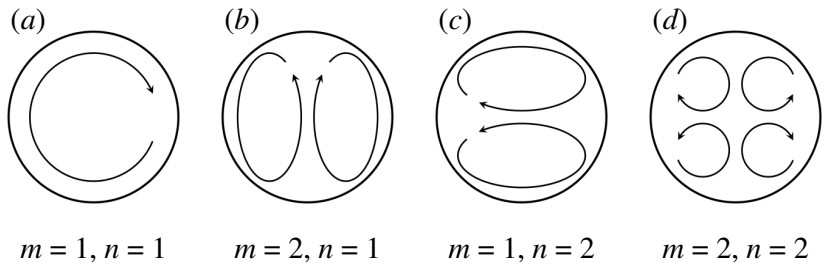}}
  \caption{Schematic illustrations of the first four Fourier modes.}
\label{fig:fourier_cartoon}
\end{figure}

Since high-order flow modes play important roles in the dynamical process of the reversal of the LSC \citep{mishra2011dynamics,xi2016higher},  we perform Fourier mode decomposition on the velocity field, which is convenient to operate and has been employed to study flow reversal mechanisms in 2-D and quasi-2-D square convection cells \citep{petschel2011statistical,chandra2011dynamics,chandra2013flow,chong2018effect,wang2018flow,chen2019emergence}.
Specifically, the instantaneous velocity field $(u, v)$ is projected onto the Fourier basis $(\hat{u}^{m,n},\hat{v}^{m,n})$ as
\begin{equation}
u(x,y,t)=\sum_{m,n}A_{x}^{m,n}(t)\hat{u}^{m,n}(x,y)
\end{equation}
\begin{equation}
v(x,y,t)=\sum_{m,n}A_{y}^{m,n}(t)\hat{v}^{m,n}(x,y)
\end{equation}
Here, the Fourier basis $(\hat{u}^{m,n},\hat{v}^{m,n})$ is chosen as
\begin{equation}
\hat{u}^{m,n}(x,y)=2\sin(m\pi x)\cos(n\pi y)
\end{equation}
\begin{equation}
\hat{v}^{m,n}(x,y)=-2\cos(m\pi x)\sin(n\pi y)
\end{equation}
The instantaneous amplitude of the Fourier mode is then calculated as
\begin{equation}
A_{x}^{m,n}(t)=\langle u(x,y,t),\hat{u}^{m,n}(x,y) \rangle=
\sum_{i}\sum_{j}u(x_{i},y_{i},t)\hat{u}^{m,n}(x_{i},y_{i})
\end{equation}
\begin{equation}
A_{y}^{m,n}(t)=\langle v(x,y,t),\hat{v}^{m,n}(x,y) \rangle=
\sum_{i}\sum_{j}v(x_{i},y_{i},t)\hat{v}^{m,n}(x_{i},y_{i})
\end{equation}
where $\langle u, \hat{u} \rangle$ and $\langle v, \hat{v} \rangle$ denote the inner product of $u$ and $\hat{u}$, and $v$ and $\hat{v}$, respectively.
This Fourier mode decomposition approach was originally designed to study flow behaviours in a square domain.
To extend this approach to the circular cell, we still choose the square domain for computing the Fourier basis but set the velocity vectors that fall outside the circular cell to be zero.
It should be noted that the Fourier basis described by Eqs. (3.3) and (3.4) may not be very accurate for a circular cell, since they were originally designed for a square cell.
However, as the Fourier basis mainly captures flows in the bulk region, and the cell corners have minor effects on the Fourier mode analysis \citep{chen2020reduced}, the Fourier basis used here indeed captures the main structure of the bulk flow.
Actually, even in a square cell, the Fourier basis functions do not satisfy the no-slip boundary condition, yet they capture well the convective flow profile \citep{chandra2011dynamics}.
The energy in each Fourier mode is calculated as $E^{m,n}(t)=\sqrt{[A_{x}^{m,n}(t)]^{2}+[A_{y}^{m,n}(t)]^{2}}$.
The $(m, n)$ Fourier mode corresponds to a flow structure with $m$ rolls in the $x$ direction and $n$ rolls in the $y$ direction, as illustrated in figure \ref{fig:fourier_cartoon}.
In the following, we will consider $m$ and $n = 1, 2, 3$, namely the first nine Fourier modes.

\begin{figure}
  \centerline{\includegraphics[width=11cm]{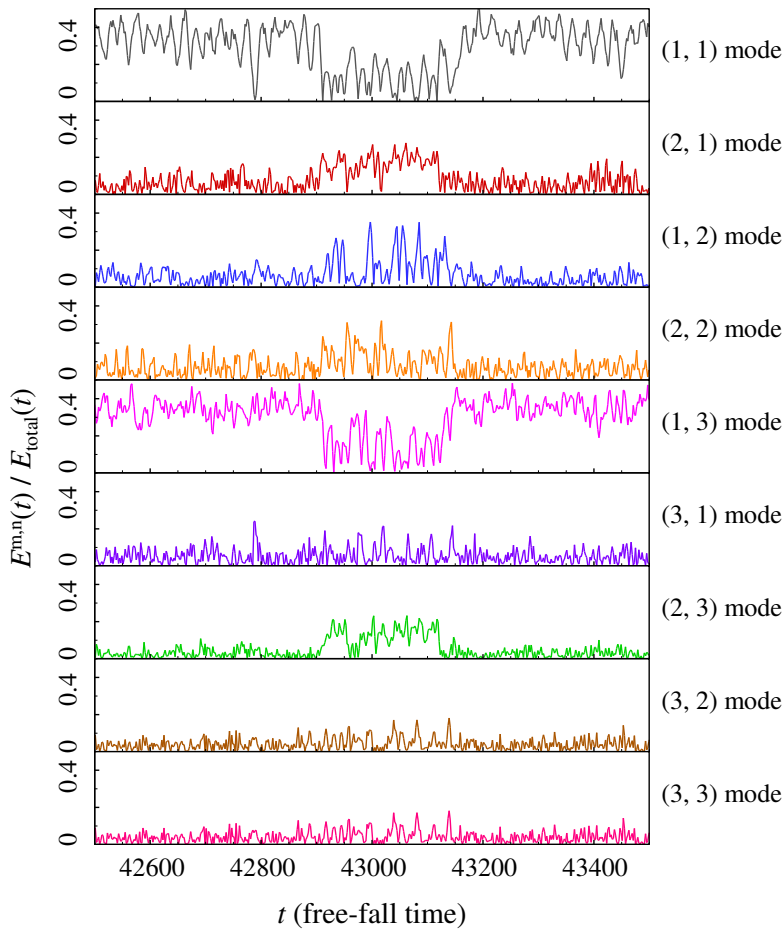}}
  \caption{Time evolution of the energy contained in the first nine Fourier modes during one reversal of the LSC for $Ra = 10^{7}$ and $Pr = 4.3$.}
\label{fig:mode_trace}
\end{figure}

The time evolution of energy contained in each Fourier mode during a flow reversal from $t = 42\ 500 \ t_f$ to $43\ 500 \ t_f$ (the same dataset as in figure \ref{fig:three_states}d) is plotted in figure \ref{fig:mode_trace}.
Here, we normalize energy of the $(m, n)$ mode $E^{m,n}$ by the total energy $E_{total}(t)=\sum E^{m,n}(t)$.
According to the time segments of the dimensionless angular momentum shown in figure \ref{fig:three_states}(d), the flow is in the stable LSC during $42\ 500 \ t_{f} \le t \le 42\ 900 \ t_{f}$ and $43\ 100 \ t_{f} \le t \le 43\ 500 \ t_{f}$.
We can see from figure \ref{fig:mode_trace} that, during this period, the dominant Fourier modes are the (1, 1) and (1, 3) modes since they account for over 20\% of the total energy.
The prevailing of the (1, 1) and the (1, 3) modes is consistent with the fact that the flow is in the shape of a primary roll in the middle, as well as secondary rolls near the top and bottom circular walls.
Between the two successive circulation states, the energies of the (2, 1), (1, 2), (2, 2) and (2, 3) modes grow, and these higher-order flow modes might correspond to the massive plume eruptions in the chaotic state.

\begin{figure}
  \centerline{\includegraphics[width=11cm]{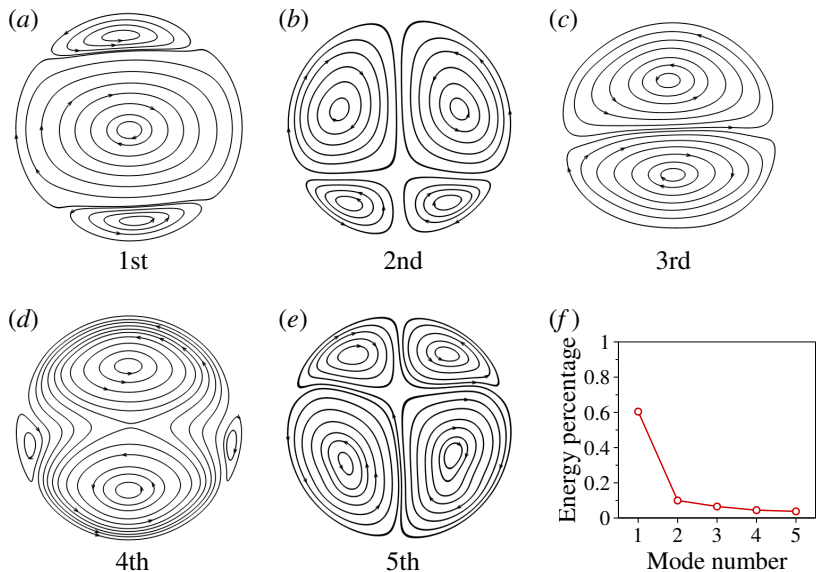}}
  \caption{(\textit{a}-\textit{e}) The first five most energetic POD modes and  (\textit{f}) the energy contained in each mode, for $Ra = 10^{7}$ and $Pr = 4.3$.}
\label{fig:POD}
\end{figure}

To check the validation of the above described Fourier mode decomposition approach for the circular cell, we perform proper orthogonal decomposition (POD) of the flow during a reversal, which does not require prior knowledge of the geometry of the flow domain \citep{berkooz1993proper}.
The POD analysis has been employed to study flow reversal mechanisms in both 2-D and three-dimensional convection cells \citep{podvin2015large,podvin2017precursor,castillo2019cessation,soucasse2019proper}.
Specifically, the flow variable $\mathbf{U}(\mathbf{x},t)$ is decomposed as a superposition of empirical orthogonal eigenfunctions $\phi_{i}(\mathbf{x})$ and their amplitudes $a_{i}(t)$ as
\begin{equation}
\mathbf{U}(\mathbf{x},t) =\sum_{i=1}^{\infty} a_{i}(t)\phi_{i}(\mathbf{x})
\end{equation}
The eigenfunctions $\phi_{i}(\mathbf{x})$ are solutions of the eigenvalue problem
\begin{equation}
\int_{\Omega}\left[\frac{1}{N}\sum_{k=1}^{N}\mathbf{U}(\mathbf{x},t_{k})\mathbf{U}(\mathbf{x}',t_{k}) \right]\phi_{i}(\mathbf{x}')d\mathbf{x}'=\lambda_{i}\phi_{i}(\mathbf{x})
\end{equation}
where $\Omega$ is the spatial domain and $N$ is the total number of snapshots.
If the empirical eigenfunctions are normalized, we have $\langle a_{i}(t) a_{j}(t) \rangle_{t}=\delta_{ij}\lambda_{i}$, where $\delta_{ij}$ is the Kroneker symbol and $\lambda_{i}$ is the energy of the $i$th POD mode.
Figure \ref{fig:POD} (\textit{a}-\textit{e}) shows the first five most energetic POD modes, which were obtained on a dataset of 40 000 snapshots spanning $10 \ 000 \ t_{f} \le t \le 50 \ 000 \ t_{f}$ for $Ra = 10^{7}$ and $Pr = 4.3$.
From the energy contained in each mode (shown in figure \ref{fig:POD}\textit{f}), we can see that the first mode accounts for over 60\% of the total energy, while each of the other higher-order modes accounts for less than 10\% of the total energy.
The most energetic POD mode consists of a primary roll in the cell centre and two secondary rolls near the top and bottom circular walls, which is similar to the mean field shown in figure \ref{fig:meanfield_Ra}.
The superposition of the second and the fifth most energetic POD modes is associated with a quadrupole flow, which corresponds to the (2, 2) Fourier mode.
The superposition of the third and the fourth most energetic POD modes consists of two large rolls superposed in the vertical direction, and it corresponds to the (1, 2) Fourier mode.
In addition, we compare the energy contained in the POD modes and the Fourier modes for the same flow dataset, as shown in table \ref{tab:PODenergy}.
We can see that the energies contained in the first five POD modes are consistent with those in (1, 1) + (1, 3), (2, 2) and (1, 2) Fourier modes, respectively.
Thus, the one-to-one correspondence between the POD mode and the Fourier mode suggests the validity of the Fourier mode decomposition in the circular cell.
In the following, we will focus on the Fourier mode decomposition to analyse the high-order flow modes, in which the same basis functions can be adopted for different $Ra$ and $Pr$ flows (in contrast to POD analysis, where the rankings of the first several major flow modes might be different for different $Ra$ and $Pr$).

\begin{table}
  \begin{center}
\def~{\hphantom{0}}
  \begin{tabular}{ccccccccccccccccccccccccccccccccccccccc}
  Fourier mode        & (1, 1) + (1, 3) & (2, 2)       & (1, 2)  \\ [3pt]
  (Energy percentage) & 55.5\%        & 8.1\%        & 7.3\%   \\
  POD mode            & 1st           & 2nd + 5th    & 3rd + 4th \\
  (Energy percentage) & 60.4\%        & 13.8\%       & 11.0\%
  \end{tabular}
  \caption{Comparison of the energy contained in the Fourier modes and the POD modes.}
  \label{tab:PODenergy}
  \end{center}
\end{table}

\begin{figure}
  \centerline{\includegraphics[width=11cm]{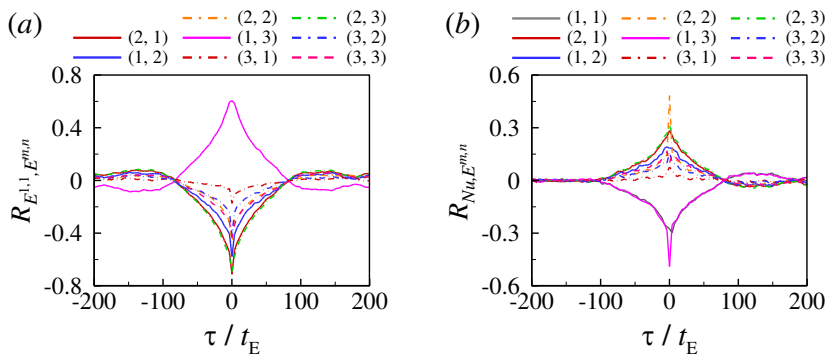}}
  \caption{Cross-correlation functions (\textit{a}) between $E^{1,1}$  and $E^{m,n}$ and (\textit{b}) between instantaneous $Nu$ and $E^{m,n}$ as a function of lag time $\tau/t_{E}$  for $Ra = 10^{7}$ and $Pr = 4.3$.}
\label{fig:correlation}
\end{figure}

To study the relationship between different Fourier modes, we calculate the cross-correlation function between the energy contained in the (1, 1) mode and the other higher-order modes as
\begin{equation}
R_{E^{1,1},E^{m,n}}(\tau)=\frac{\langle [E^{1,1}(t+\tau)-\langle E^{1,1} \rangle][E^{m,n}(t)-\langle E^{m,n} \rangle]\rangle}{\sigma_{E^{1,1}} \sigma_{E^{m,n}}},
\end{equation}
where $\sigma_{E^{m,n}}$ is the standard deviation of the energy of the $(m, n)$ mode $E^{m,n}$.
As shown in figure \ref{fig:correlation}(a),  $R_{E^{1,1},E^{1,3}}$ shows a positive peak near $\tau/t_{E}=0$, implying that the strengths of the (1, 1) and (1, 3) flow modes are positively correlated;
meanwhile, the strengths of other higher-order modes are negatively correlated with $E^{1,1}$.
The results from this statistical analysis are consistent with what we have learned from the instantaneous example (see figure \ref{fig:mode_trace}): when the (1, 1) mode becomes weaker, the (1, 3) mode also becomes weaker, and vice versa.
It is also noted that the correlations between the (1, 1) mode and the (2, 2), (2, 3) and (2, 1) modes all have negative peaks near $\tau/t_{E}=0$, indicating these higher-order modes themselves are positively correlated.
In addition, we study the relationship between the heat transfer efficiency and the energy contained in different modes by calculating their cross-correlation functions as
\begin{equation}
R_{Nu,E^{m,n}}(\tau)=\frac{\langle [Nu(t+\tau)-\langle Nu \rangle][E^{m,n}(t)-\langle E^{m,n} \rangle]\rangle} {\sigma_{Nu} \sigma_{E^{m,n}}},
\end{equation}
where $\sigma_{Nu}$ and $\sigma_{E^{m,n}}$ are the standard deviation of $Nu$ and the energy of the $(m, n)$ mode $E^{m,n}$, respectively.
As shown in figure \ref{fig:correlation}(b), $Nu$ is negatively correlated with $E^{1, 1}$ and $E^{1, 3}$, which means that the (1, 1) and (1, 3) flow modes are associated with smaller heat transfer efficiency on average;
meanwhile, the other higher-order modes, including the (2, 2), (2, 3) and (2, 1) modes, are positively correlated with $Nu$.
These results are consistent with results from the conditional averaged Nusselt number shown in figure \ref{fig:Nu_diff_state}(a), i.e. $Nu$ is smaller when the flow is in the circulation states (corresponding to the (1, 1) and (1, 3) flow modes).

\begin{figure}
  \centerline{\includegraphics[width=12cm]{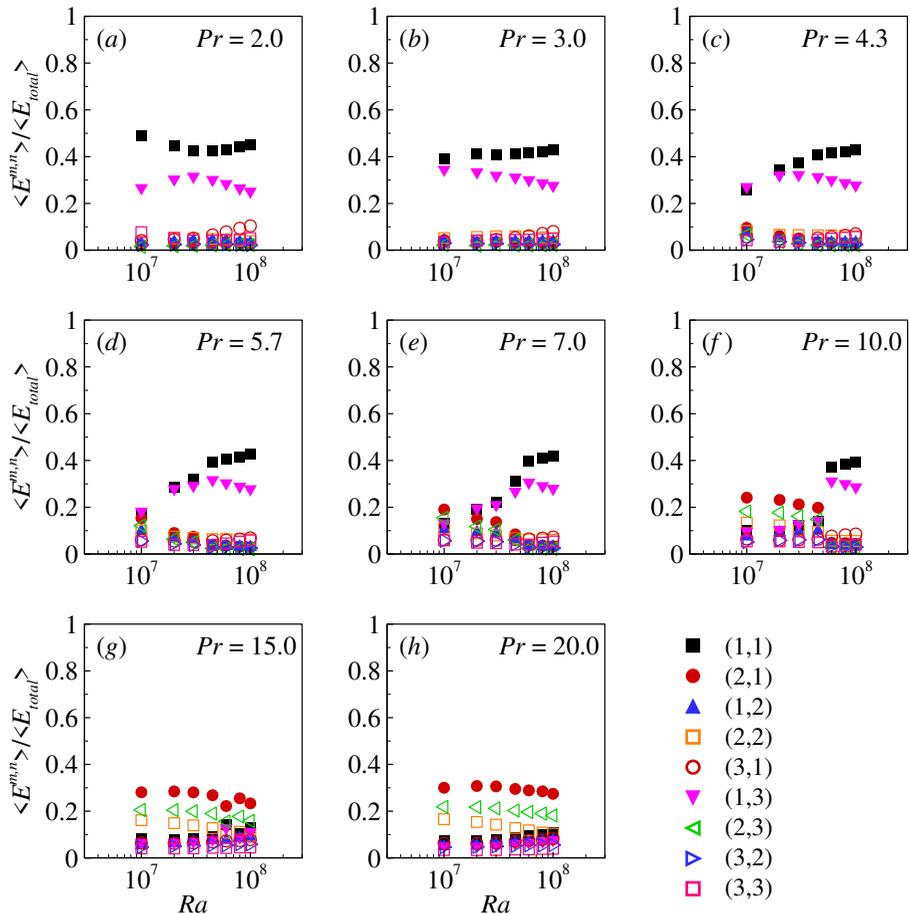}}
  \caption{Time-averaged energy contained in each of the first nine Fourier modes as functions of $Ra$ for (\textit{a}) $Pr = 2.0$, (\textit{b}) $Pr = 3.0$, (\textit{c}) $Pr = 4.3$, (\textit{d}) $Pr = 5.7$, (\textit{e}) $Pr = 7.0$, (\textit{f}) $Pr = 10.0$, (\textit{g}) $Pr = 15.0$ and (\textit{h}) $Pr = 20.0$.}
\label{fig:mode_Ra}
\end{figure}

We further calculate the time-averaged energies contained in each Fourier mode $\langle E^{m,n} \rangle$ as a function of $Ra$ for various $Pr$, and normalize the values with the time-averaged total energies $\langle E_{total} \rangle$.
From figure \ref{fig:mode_Ra} we can see that, at lower $Pr$ (e.g.  $Pr = 2.0$ and $Pr=3.0$), the (1, 1) and (1, 3) modes dominate and they together contain more than 70\% of the total flow energy, resulting in a stable LSC and no (or rare) flow reversal events.
At higher $Pr$ (e.g.  $Pr = 20.0$), both the (1, 1) and (1, 3) modes are very weak, and thus a stable LSC does not exist.
At intermediate $Pr$, the (1, 1) and (1, 3) modes are weaker at lower $Ra$, and thus the LSC can be easily broken, further leading to its reversal;
with increasing $Ra$, the (1, 1) and (1, 3) modes become stronger, implying that the LSC is stable.
We also note the $Ra$ and $Pr$ dependencies of the time-averaged energies are similar to those of the residence time shown in figure \ref{fig:residence_time}, mainly because the (1, 1) and (1, 3) modes are dominant in the circulation states.

\subsection{Phase diagram of the flow states and reversal in $Ra$-$Pr$ plane}

\begin{figure}
  \centerline{\includegraphics[width=11cm]{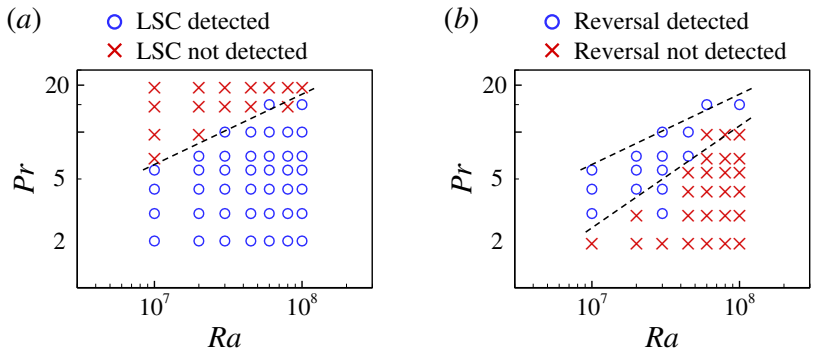}}
  \caption{Phase diagram of (\textit{a}) the flow states and (\textit{b}) the flow reversals in the $Ra$-$Pr$ plane. The grey dashed lines represent the estimated boundaries between the different regimes. The upper boundary in (\textit{b}) is the same as the boundary in (\textit{a}).}
\label{fig:phase_diagram}
\end{figure}

With long-time simulations in a wide range of $Ra$ and $Pr$ parameter spaces, we can then obtain the phase diagram of whether the flow is in the form of an LSC and whether the reversal of the LSC could occur, as shown in figure \ref{fig:phase_diagram}.
The criteria to determine whether an LSC exists are based on checking the value of dimensionless angular momentum.
If $L/L_{0}$ reaches a plateau well above (below) zero with positive (negative) fluctuations, it corresponds to an LSC organized in counterclockwise (clockwise) direction.
The criteria to determine whether the LSC reversals occur are described in Section \ref{sec:tristable}.
The phase diagram in the $Ra-Pr$ plane can be understood in terms of competition between the thermal and viscous diffusions.
For high $Pr$ or low $Ra$ (corresponding to the top-left region of the phase diagram in figure \ref{fig:phase_diagram}a), the flow is in the chaotic state, and the LSC is not formed, let alone the reversal of the LSC; for low $Pr$ or high $Ra$, the flow is in the circulation states, and there is only stable LSC (see the bottom-right region of the phase diagram in figure \ref{fig:phase_diagram}b). From figure \ref{fig:phase_diagram}b, we can see that the LSC reversal only occurs at a limited $Ra$ and $Pr$ range.
We note that the phase diagram shown here is very similar to that for the 2-D/quasi-2-D square cell in \cite{sugiyama2010flow} where the phase diagram is understood based on the growth of the corner rolls.
And the reason that there is no reversal occurring for very small $Pr$ or very high $Pr$ is that in those cases the building up of the corner rolls is suppressed.
As in the circular cell the LSC is still not a single roll but with two secondary rolls, the reversal must be understood in terms of the interaction between the main roll and the corner rolls.
Thus the argument used in \cite{sugiyama2010flow} for the 2-D/quasi-2-D square cell could be directly borrowed to understand the reversals in the circular cell, at least in the parameter range of the current study.
The flow in the circular cell will probably evolve into a single roll without the secondary rolls when $Ra$ becomes very large.
In that regime, the reversal then should be understood solely in terms of the dynamics of the single roll itself.

\section{Conclusions}\label{sec:conclusions}

In this work, we have performed long-time direct numerical simulations of turbulent thermal convection in 2-D circular cells.
We observed three meta-stable flow states in the convection of a 2-D circular cell, namely two circulation states (both clockwise and counterclockwise) and a chaotic state characterized with incoherent plume motions.
The chaotic state is a manifestation of the fact that it is difficult for the plumes to organize into the coherent LSC.
We found that the chaotic state dominates the flow for high $Pr$, while the circulations states dominate the flow for low $Pr$.
For intermediate $Pr$, the circulation states become stronger while the chaotic state becomes weaker with increasing $Ra$.
We also found that the internal flow states manifested themselves into global properties such as Nusselt number and Reynolds number: $Nu$ is up to $8\%$ larger in the chaotic state than that in the circulation states and $Re$ is up to $17\%$ smaller in the chaotic state than that in the circulation states.
Intuitively one would imagine that more coherent and stronger flow would have higher heat transfer efficiency, while the results shown here, i.e. the stronger and more coherent circulation state inducing lower $Nu$, are counterintuitive and await further study.

The original objective of adopting the circular cell was to make the geometry of the cell to perfectly fit the circulatory flow, i.e. the LSC, and eliminate the corner rolls and their effects on the dynamics such as the reversal of the LSC.
While it is found that in the circular cell it is not the case that one single LSC occupies the whole cell, at least in the parameter range of the current study, and instead the circulation states of the flow consist of one primary roll in the middle and two secondary rolls near the top and bottom circular walls.
The top and bottom secondary rolls in the circular cell and the two corner rolls in the square cell are from the same origin, namely the interaction between the mean flow and the plumes erupted from edges (corners) of the top and bottom circular (flat) walls.
The primary roll becomes stronger and larger, while the two secondary rolls become weaker and smaller, with increasing $Ra$.
When the flow transits from the circulation state to the chaotic state, the secondary rolls play important roles.
They grow in size and energy, squeeze and finally break up the primary roll.
The reversal of the LSC is accompanied by the growth of top and bottom secondary rolls, the breaking of the primary roll and the connection of the secondary rolls.

Finally, we mapped out the phase diagram of whether the flow is organized as an LSC and whether reversal could occur.
In the range $10^{7} \le Ra \le 10^{8}$ and $2.0 \le Pr \le 20.0$, it is found the flow is in an organized LSC when $Ra$ is large and $Pr$ is small.
The reversal of the LSC could only occur in a limited $Pr$ and $Ra$ range.
As in the parameter range of the current study the circulation state of the flow consists of the main roll and the secondary rolls, the reversal of the LSC should not be understood solely in terms of the dynamics of the main roll itself, but a combined effect of the interaction between the main roll and the secondary rolls, and the stability of the main roll itself.

\section{Acknowledgments}
This work was supported by the National Natural Science Foundation of China (NSFC) through Grant nos. 11902268 and 11772259, the Fundamental Research Funds for the Central Universities of China (nos. D5000200570 and 3102019PJ002) and the 111 project of China (no. B17037).

\section{Declaration of interests}
The authors report no conflict of interest.

\section{Supplementary movies}
Supplementary movies are available at \url{https://doi.org/10.1017/jfm.2020.964}.

\bibliographystyle{jfm}
\bibliography{myBib}

\end{document}